\documentstyle[twoside,fleqn,espcrc2,epsfig]{article}

\def\beq{\begin{equation}}
\def\eeq{\end{equation}}
\def\ni{\noindent}

\newcommand{\AmS}{{\protect\the\textfont2
  A\kern-.1667em\lower.5ex\hbox{M}\kern-.125emS}}

\title{Results from the ARGO-YBJ Test Experiment}

\author{The ARGO-YBJ Collaboration\thanks{
C. Bacci$^1$, K.Z. Bao$^2$, F. Barone$^3$, B. Bartoli$^3$,
P. Bernardini$^4$, S. Bussino$^1$, E. Calloni$^3$, 
B.Y. Cao$^5$, R. Cardarelli$^6$, S. Catalanotti$^3$, A. Cavaliere$^6$, 
S. Cavaliere$^3$, F. Cesaroni$^4$,
P. Creti$^4$, Danzengluobu$^7$, B. D'Ettorre Piazzoli$^3$, M. De Vincenzi$^1$, 
T. Di Girolamo$^3$, G. Di Sciascio$^3$, Z.Y. Feng$^8$, Y. Fu$^5$, X.Y. Gao$^9$, 
Q.X. Geng$^9$, H.W. Guo$^7$, H.H. He$^{10}$, M. He$^7$, Q. Huang$^8$,
M. Iacovacci$^3$, N. Iucci$^1$, H.Y. Jai$^8$, F.M. Kong$^5$, H.H. Kuang$^{10}$, 
Labaciren$^7$, B. Li$^2$, J.Y. Li$^5$, Z.Q. Liu$^9$, H. Lu$^{10}$, 
X.H. Ma$^{10}$, G. Mancarella$^4$, S.M. Mari$^{11}$, G. Marsella$^4$, 
D. Martello$^4$, D.M Mei$^7$, X.R. Meng$^7$, L. Milano$^3$, A. Morselli$^6$, 
J. Mu$^9$, M. Panareo$^4$, Z.R. Peng$^{10}$, P. Pistilli$^1$, 
R. Santonico$^6$, P.R. Shen$^{10}$, C. Stanescu$^1$, J. Su$^{10}$, 
L.R. Sun$^2$, S.C. Sun$^2$, A. Surdo$^4$, Y.H. Tan$^{10}$, 
S. Vernetto$^{12}$, C.R. Wang$^5$, F. Wang$^{10}$, H.Y. Wang$^{10}$,
Y.N. Wei$^2$, H.T. Yang$^9$, Q.K. Yao$^2$, G.C. Yu$^8$, X.D. Yue$^2$,
A.F. Yuan$^7$, H.M. Zhang$^{10}$, J.L. Zhang$^{10}$,
N.J. Zhang$^5$, T.J. Zhang$^9$, X.Y. Zhang$^5$,
Zhaxisangzhu$^7$, Zhaxiciren$^7$, Q.Q. Zhu$^{10}$.
\it {
$^1$ INFN and Dipartimento di Fisica dell'Universit\`a di Roma Tre, Italy,
$^2$ Zhenghou University, Henan, China,
$^3$ INFN and Dipartimento di Fisica dell'Universit\`a di Napoli, Italy,
$^4$ INFN and Dipartimento di Fisica dell'Universit\`a di Lecce, Italy,
$^5$ Shangdong University, Jinan, China,
$^6$ INFN and Dipartimento di Fisica dell'Universit\`a di Roma "Tor Vergata", 
Italy,
$^7$ Tibet University, Lhasa, China,
$^8$ South West Jiaotong University, Chengdu, China,
$^9$ Yunnan University, Kunming, China,
$^{10}$ IHEP, Beijing, China,
$^{11}$ Universit\'a della Basilicata, Potenza, Italy,
$^{12}$ Istituto di Cosmo-Geofisica del CNR and INFN, Torino, Italy}}\\[0.3cm]
Presented by M. Iacovacci
\address{Dipartimento di Fisica dell'Universit\`a di Napoli and 
I.N.F.N., Napoli, Italy}}

\begin{document}

\begin{abstract}
An RPC carpet covering $\sim$ 10$^4$ m$^2$ (ARGO-YBJ experiment) will be
installed in the YangBaJing Laboratory (Tibet, P.R. China) at an altitude of
4300 m a.s.l.. A test-module of $\sim$ 50 m$^2$ has been put in operation
in this laboratory and about 10$^6$ air shower events have been collected.
The carpet capability of reconstructing the shower features is presented.
\end{abstract}

\maketitle

\section{INTRODUCTION}

The ARGO-YBJ experiment is under way over the next few years at Yangbajing
High Altitude Cosmic Ray Laboratory (4300 m a.s.l., 606 $g/cm^2$), 90 km
North to Lhasa (Tibet, P.R. China).
The aim of the experiment is the study of cosmic rays, mainly
cosmic $\gamma$-radiation, at an energy threshold of $\sim 100$ $GeV$, by
means of the detection of small size air showers at high altitude.
The apparatus consists of a full coverage detector of dimension
$\sim 71\times 74\>m^2$ realized with a single layer of Resistive Plate
Counters (RPCs). The area surrounding the central detector core, up to $\sim
100\times 100\>m^2$, consists of a guard ring partially ($\sim 50\> \%$)
instrumented with RPCs.
These outer detector improves the apparatus performance, enlarging the
fiducial area, for the detection of showers with the core outside the
full coverage carpet.
A lead converter $0.5$ $cm$ thick will cover uniformly the RPC plane in order
to increase the number of charged particles by conversion of shower photons
and to reduce the time spread of the shower front.
The site location (longitude $90^{\circ}$ 31' 50'' E, latitude
$30^{\circ}$ 06' 38'' N) permits the monitoring of the Northern hemisphere in
the declination band $-10^{\circ}<\delta <70^{\circ}$. 

Such a
detector, performing a continuous high sensitivity sky survey, complements
the narrow field of view air Cerenkov telescopes allowing to bridge
the GeV and TeV energy regions and to face a wide range of fundamental issues
in Cosmic Ray and Astroparticle Physics including $\gamma$-ray astronomy,
GRBs physics and the measurement of the $\overline{p}/p$ at TeV energies 
\cite{abbr96}.
Detector assembling will start late in 2000 and data taking with the first
$\sim$ 750 $m^2$ of RPCs in 2001.

In order to investigate both the RPCs performance at 4300 $m$ a.s.l. and
the capability of the detector to sample the shower front of atmospheric
cascades, during 1998 a full coverage carpet of
$\sim 50\>m^2$ has been put in operation in the YangBaJing (YBJ) Laboratory.
Results concerning both the RPCs behaviour and the features of the showers 
imaged by the carpet are presented. 

\section{THE EXPERIMENTAL SET-UP}

The set-up of the detector is an array of $3\times 5$ chambers of area
$280 \times 112$ $cm^2$ each, covering a total area of $\sim 8.6 \times
6.0$ $m^2$. The active area of $\sim 45.8$
$m^2$, accounting for a dead area due to a $7$ $mm$ frame closing
the chamber edge, corresponds to a $\sim 89 \%$ coverage. The RPCs,
with a $2$ $mm$ gas gap, are built with bakelite electrode plates of
volume resistivity in the range ($0.5\div 1$) $10^{12} \Omega
\cdot cm$. 
The RPC signals are picked up by means of aluminum strips $3.3$ $cm$
wide and $56$ $cm$ long which are glued on a $0.2$ $mm$ thick film of
Poly-Ethylen-Tereftalate (PET). At the edge of the
detector the strips are connected to the front-end electronics and
terminated with 50 $\Omega$ resistors. A grounded aluminum foil is used to
shield the bottom face of the RPC and an extra PET foil, on top of
the aluminum, is used as a further high voltage insulator.
The front-end circuit contains 16 discriminators, with about $50$ $mV$
voltage threshold, and provides a FAST-OR signal with the same
input-to-output delay ($10$ $ns$) for all the channels. This signal is used
for time measurements and trigger purposes in the present test.
The 16 strips connected to the same front-end board are logically
organized in a pad of $56 \times 56$ $cm^{2}$ area. Each RPC is
therefore subdivided in 10 pads which work like independent
functional units. The pads are the basic elements ("pixel") which define the
space-time pattern of the shower; they give indeed the position
and the time of each detected hit. At any trigger occurrence the
times of all the pads are read-out by means of multihit TDCs of $1$
$ns$ time bin, operated in common STOP mode.
Since the pad signal is shaped to 1.5 $\mu s$, only the time profile of the 
earliest particles hitting the pads is imaged.
The set-up was completed with a small telescope consisting of 3 RPCs of
$50 \times 50$ $cm^{2}$ area with 16 pick-up
strips $3$ $cm$ wide connected to front-end electronics board
similar to the ones used in the carpet. The 3 RPCs were overlapped
one on the other and the triple coincidence of their FAST-OR
signals was used to define a cosmic ray crossing the telescope.

\section{RPC PERFORMANCE}

The RPCs were operated in streamer mode
as foreseen for the final experiment. This mode delivers
large amplitude saturated signals and is less sensitive than
the avalanche or proportional mode to electromagnetic
noise, to changes in the environment conditions and to mechanical
deformations of the detector. 
\begin{figure}[htb]
\mbox{\epsfig{file=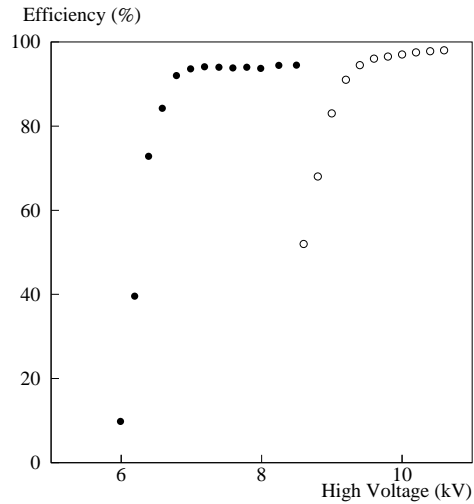,height=7.cm,width=7.cm}}
\caption{Detection efficiency $vs$ operating voltage for one of the
carpet RPCs ($\bullet$). The same curve for a 2 $mm$ gap RPC operating at
sea level is also reported ($\circ$) for comparison.}
\label{effic}
\end{figure}
Three gas mixtures were tested which used the same components, Argon,
Isobutane and TetraFluoroEthane, in different proportions:
TFE/Ar/i-But = 45/45/10; 60/27/13 and 75/15/10.
An higher TFE concentration in place of the Ar concentration
increases the primary ionization thus compensating
for the $40\%$ reduction caused by the lower gas target pressure
(600 $mbar$) and reduces the afterpulse probability.
The reduction of the Argon concentration in favour of TFE results
in a clear increase of the operating voltage as expected from the
large quenching action of TFE.
Since the higher the TFE fraction, the lower is the charge delivered in the gas
by a single streamer, we decided to operate the test carpet with the gas
mixture corresponding to the highest TFE fraction, in order 
to extend the dynamic range achievable for the analogical read-out.

Fig. \ref{effic} shows the operating efficiency for 4 ORed pads. The
efficiency was measured using cosmic ray signals defined by means of an
auxiliary telescope placed on top of the carpet.
The same curve for a $2$ $mm$ gap RPC operated at sea level
is also shown for comparison. The detection efficiency $vs$ operating voltage,
compared to the operation at $606$ $mbar$ pressure in YBJ,
shows an increase of $\sim 2.5$ $kV$ in operating voltage.
The effect
of small changes in temperature $T$ and pressure $P$ on the operating
voltage can be accounted for by rescaling the applied
voltage $V_a$ according to the relationship:
\beq
  V=V_{a}\frac{P_{0}}{P}\cdot \frac{T}{T_{0}}
\eeq
\ni where $P_0$ and $T_0$ are arbitrary standard values, e.g. $1010$ $mbar$ and
$293$ $K$ respectively for a sea level laboratory. This formula predicts,
starting from the YBJ data, an operating voltage at sea level
which is considerably smaller than the experimental one.
However, a good consistency is recovered by assuming that, in the ideal gas
approximation, the parameter which fixes the operating
voltage is given by {\em gap $\cdot$ pressure / temperature}.
Measurements on RPCs with different gas gap size justify this
assumption.
\begin{figure}[htb]
\mbox{\epsfig{file=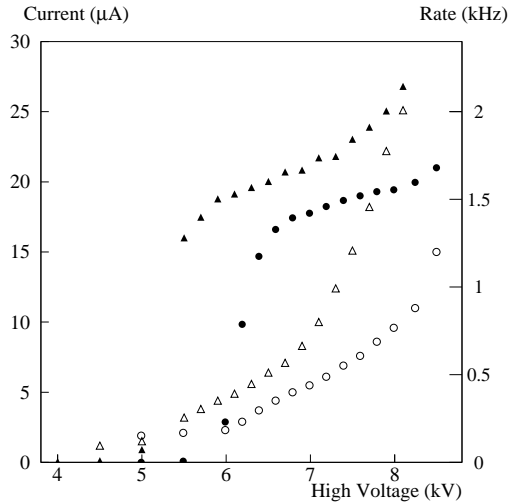,height=7.cm,width=7.cm}}
\caption{Counting rate (full symbols) and operating current (open symbols) 
vs Voltage of one RPC of the carpet. Results are presented for two 
different TFE percentages ($45\%$:triangles and $75\%$:circles). The rate shown 
refers to 4 of the 10 ORed RPC pads.}
\label{countrate}
\end{figure}

Fig. \ref{effic} also shows that the plateau efficiency measured at
YBJ is $3\div 4\%$ lower than at the sea level. Although a lower
efficiency is expected from the smaller number of primary clusters
at the YBJ pressure, we attribute most of the difference to the underestimation
of the YBJ efficiency. At YBJ altitude, 
indeed, the ratio of the cosmic radiation electromagnetic to muon
component is $\sim 4$ times larger than at sea level. A spatial tracking
with redefinition of the carpet track downstream would
eliminate the contamination from soft particles, giving  a more
accurate and higher efficiency. On the other hand the lower
efficiency could hardly be explained with the gas lower density.
In fact, the number of primary clusters in the YBJ test, estimated around
$9$, is the same as in the case of some gas, e.g.
Ar/i-But/CF3Br = 60/37/3, that was frequently used at sea level with
efficiency of $\sim 97\div 98\%$. 

The counting rate of 4 ORed pads, together with the RPC current, are 
reported in Fig. \ref{countrate} $vs$ the operating voltage. The results for 
a percentage of $45\%$ of TFE are also reported for comparison. 
A rather flat singles counting rate plateau is observed at a level
of $\sim 400$ $Hz$ for a single pad of area $56\times 56$ $cm^{2}$.
\begin{figure}[htb]
\mbox{\epsfig{file=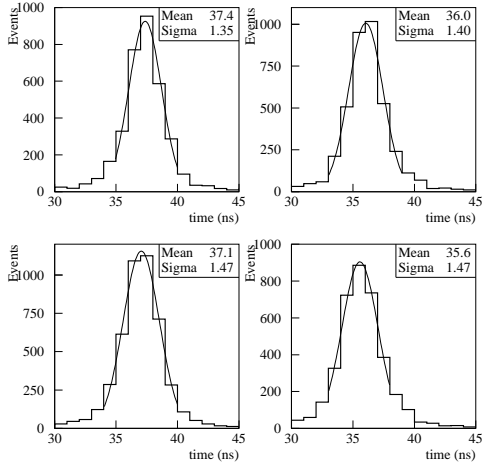,height=7.cm,width=7.cm}}
\caption{Time jitter distribution of 4 pads of the carpet. The telescope 
RPC2 signal is used as common stop.}
\label{timej}
\end{figure}

The time jitter distribution of the
pad signals was obtained by measuring the delay of the FAST-OR signal
with respect to RPC2 (the RPC in the central position
of the trigger telescope) by means of a TDC
with 1 $ns$ time bin. This distribution is shown in Fig. \ref{timej} for
four pads. The average of the standard deviations is $1.42$ $ns$
corresponding to a resolution of $\sim 1$ $ns$ for the single RPC if we
account for the fact that the distributions show the combined jitter of two
detectors. A more detailed presentation of these results is reported in 
\cite{bacci99}. 

\section{DATA ANALYSIS OF SHOWER EVENTS}

\begin{figure}[htb]
\mbox{\epsfig{file=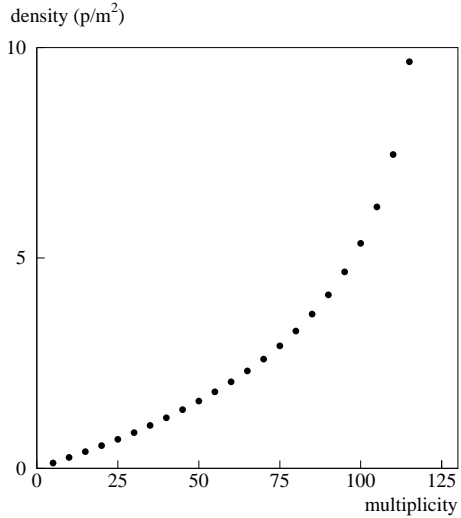,height=7.cm,width=7.cm}}
\caption{The particle density as a function of the measured multiplicity.}
\label{densit}
\end{figure}

Data were taken either with or without a 0.5 $cm$ layer of lead
on the whole carpet in order to investigate the converter effect on
multiplicity and angular resolution.
A trigger based on pad multiplicity has been used to collect
$\sim 10^6$ shower events in April-May 1998.
\begin{figure}[htb]
\mbox{\epsfig{file=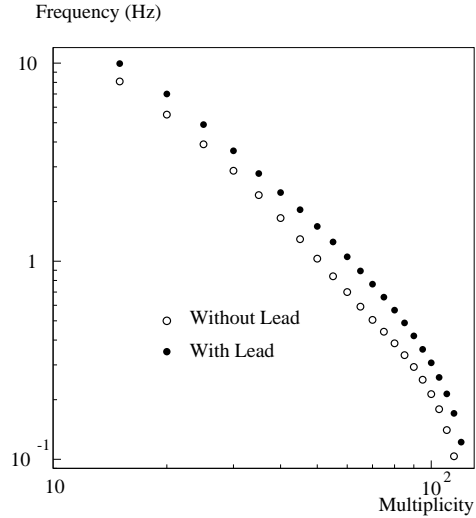,height=7.cm,width=7.cm}}
\caption{The integral rate as a function of the pad multiplicity.}
\label{freq}
\end{figure}
The integral rate as a function of the pad multiplicity is shown in 
Fig. \ref{freq} for showers before and after the lead was installed.
A comparison at fixed rate indicates an increase of pad multiplicity due
to the effect of the lead of $\sim 15\div 20\%$, as expected according to our
simulations.
Since in this test the strip read-out is not effective,
the actual multiplicity is understimed for particle densities higher than 
$\sim 1$ $m^{-2}$.
Data have been corrected to obtain the average particle density as a function
of the recorded multiplicity (Fig. \ref{densit}).

\subsection{Event reconstruction}

Due to the reduced size of the detector, the time profile of the shower front
sampled by the carpet is expected to exhibit a planar shape.
In this approximation the expected particle arrival time is a linear
function of the position.
We performe an optimized reconstruction procedure of the shower direction
as follows:
\begin{itemize}
\item  Unweighted plane fit to hits for each event by minimization of the
function
\beq
\chi^2 = \frac{1}{c^2}\sum_i\{lx_i + my_i + nz_i - c(t_i-t_0)\}^2
\eeq
The sum includes all counters with a time signal $t_i$, $c$ is the light
velocity, $(x_i,y_i,z_i)$ are the $i$th center pad position coordinates.
The parameters of the fit are the time offset $t_0$ and the $l,m$ direction
cosines.
\item Rejection of outlying points by means of a $K\cdot \sigma$ cut and
iteration of the fit until all points verify this condition. 
If the remaining hits number is $\leq 5$ the event is rejected. 
Here $\sigma$ is the standard deviation of the distribution.
\end{itemize}
This procedure is rather fast because it makes use only of analytic
formulae. No {\em a priori} information about shower features is required.
The actual value of $K$ could depend on the features of
the reconstructed showers as well as on the experimental conditions 
(pad dimension, shaping of the signals, multihit capability, etc.). By
choosing $K=2.5$ about $8\%$ of the hits are rejected. Increasing
$K$ should increase the number of discarded hits without a significant
improvement of the $\chi^2$. 
The $\chi^2$ distribution exhibits a long tail more pronounced for low 
multiplicity events. In these high $\chi^2$ events the time hits are 
considerably spread, by far more than expected from the detector time 
resolution. They can be attributed to the sampling of a portion of the 
shower affected by large time fluctuations. Events with reduced $\chi^2>30$ 
$ns^2$ ($\sim 10\%$ of the total) have been discarded and not used in the 
following analysis. 

\subsection{The zenith angle distribution}

\begin{figure}[htb]
\mbox{\epsfig{file=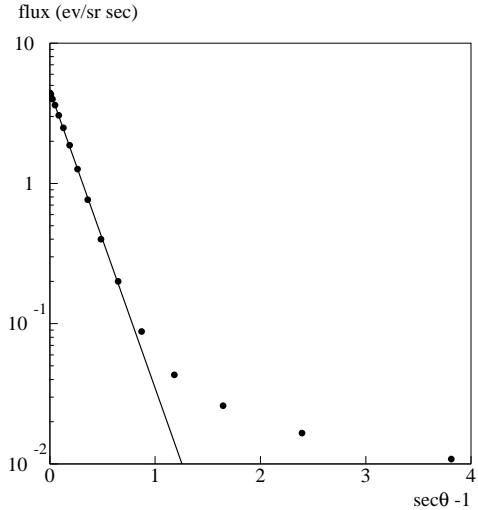,height=7.cm,width=7.cm}}
\caption{The zenith angle distribution.}
\label{zenith}
\end{figure}

Since the shower size is not measured, events have been classified
according to the observed particle density.
The zenith angle distribution of the integral density spectrum is expected
to follow an exponential behaviour
\beq
I(\geq \rho,\theta)=I(\geq \rho,0)\cdot e^{ -{x_0 \over {\Lambda_{att}} }
(sec\theta -1) }
\eeq
where $x_0$ is the vertical depth. The attenuation mean free path
$\Lambda_{att}$ of EAS is related to the absorption length $\Lambda_e$ of
EAS particles by $\Lambda_{att}=\Lambda_e/\gamma$ where $\gamma$ is the index
of the integral density spectrum. The angular distribution of events with
density greater than 3.0 particles/$m^2$ (corresponding to
$N_{hit}(\theta)$ = $76\cdot cos\theta$) is shown in Fig. \ref{zenith} .
The data can be fit out to $\sim 55^{\circ}$
with an $exp[-\alpha\cdot(sec\theta-1)]$ law. The parameter $\alpha$ is
found to be $4.88\pm 0.45$, so that $\Lambda_{att}=(124\pm 11)$ $g/cm^2$, 
in excellent agreement with previous results concerning the density spectrum 
of shower particles \cite{bennet61}. 
For angles greater than $55^{\circ}$ a deviation from this law is observed.
Misreconstructed events, showers locally produced in the walls of the
sourrounding buildings and horizontal air showers could contribute to these
large angle events.
This result is substantially unchanged for densities ranging from 1 to 5 
particles/$m^2$.

Data confirm that the
shape of the angular distribution is dominated by the physical effect of
atmospheric absorption and that no relevant instrumental effects or
selection biases have been introduced.

\subsection{The density spectrum}

\begin{figure}[htb]
\mbox{\epsfig{file=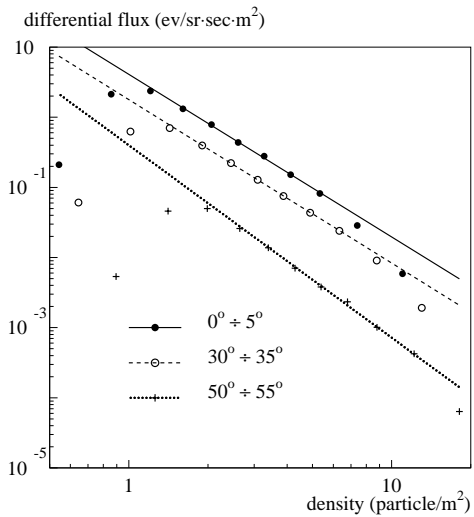,height=7.cm,width=7.cm}}
\caption{Differential density spectrum for three intervals of zenith angle.}
\label{densthet}
\end{figure}

The differential density spectrum measured in three different intervals of
zenith angles is shown in Fig. \ref{densthet} . The angular bin size is $\Delta
\theta=5^{\circ}$, so that
the  difference of atmospheric depth inside the source bin is $\leq 40$ 
$g/cm^2$ (about one radiation length) for zenith angles up to $35^{\circ}$.
A smaller bin size has been used to analyze data up to $55^{\circ}$. 
The spectra refer to the depth corresponding to the mean zenith angle of 
the showers recorded inside each interval.
The shapes of the spectra are very similar in the density range $2 \div 8$ 
particles/$m^2$ not affected by threshold or saturation effects. 

\begin{figure}[htb]
\mbox{\epsfig{file=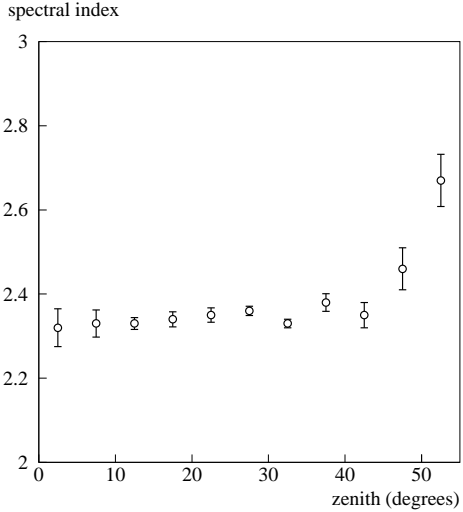,height=7.cm,width=7.cm}}
\caption{The index $\gamma_d$ of the density spectrum as a function of the 
zenith angle.}
\label{spectra}
\end{figure}

The index $\gamma_d$ of the density spectrum as a function of the zenith angle
is shown in Fig. \ref{spectra}. A weighted mean up to 
$\theta < 45^{\circ}$ gives $2.33\pm 0.03$. 
This value is fully consistent with the ones obtained in experiments at high 
altitude at comparable particle densities \cite{greisen56}.
MonteCarlo simulations show that present data concern energies ranging from 
about 1 TeV (quasi-vertical showers) to a few hundred TeV ($\theta \sim 
45^{\circ}$). Accordingly, the measured $\gamma_d$
can be considered a reasonable estimate of the slope of the density spectrum 
of showers initiated by primaries (mainly protons) of energies 
$\sim 10^{13}\div 10^{14}$ $eV$. From these
data the absorption length of EAS particles $\Lambda_e$ is found to be 
$165\pm 18$ $g/cm^2$ consistent
with results of similar measurements at small shower sizes \cite{bennet61}.
Data recorded at large zenith angles ($> 45^{\circ}$) are due to showers 
developed through atmospheric depths greater than $860$ $g/cm^2$. Thus, a 
substantial contribution is expected from showers originated by primaries 
of energies beyond the knee of the cosmic ray spectrum. This could 
explain the increasing of the slope $\gamma_d$ with zenith angle, as shown 
in Fig. \ref{spectra}. 

\subsection{Shower front thickness}

\begin{figure}[htb]
\mbox{\epsfig{file=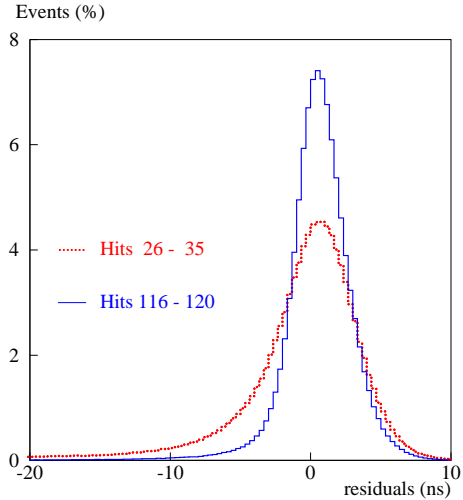,height=7.cm,width=7.cm}}
\caption{Distribution of time residuals for events with different pad
multiplicity (all channel added).}
\label{timethick}
\end{figure}

The time distribution of the shower hits with respect to the fitted plane is
shown in Fig. \ref{timethick} for two different multiplicity ranges.
Quasi-vertical showers ($\theta<15^{\circ}$) have been selected. 
The distribution of time residuals 
exhibits, as expected, a long tail due to time fluctuations and to the
curved profile of the shower front, more pronounced for low multiplicity.
The width of these distributions is related to the time thickness of the
shower front. Since the position of the shower core is not reconstructed,
the experimental result concerns a time thickness averaged on different radial
distances. 

For low multiplicity events with a mean
number of particles per pad $<1$, the spread $\sigma$ represents a
reasonable measurement of the shower thickness. For high multiplicity
events ($N_{hit}>100$), where the mean number of particles hitting one pad
is $>1$, this $\sigma$ does not express the thickness of the shower disc
without any bias. In fact, the time residual distribution is related to the
fluctuations of the first particle detected by each pad.
Taking into account the total detector resolution of $1.3$ $ns$ (RPC
intrinsic jitter, strip length, electronics time resolution) the time
jitter of the earliest particles in high multiplicity events
($>100$ hits) is estimated $\sim 1$ $ns$. 
The averaged shower front thickness of low multiplicity events (particle 
density $< 1$ $m^{-2}$) is found $\sim 4.4$ $ns$.

\subsection{Angular resolution}

\begin{figure}[htb]
\mbox{\epsfig{file=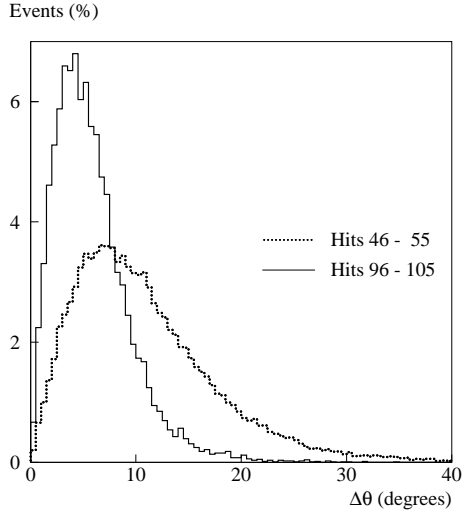,height=7.cm,width=7.cm}}
\caption{Even-odd angle difference distribution for events with
different pad multiplicity.}
\label{evenodd}
\end{figure}

The angular resolution of the carpet has been estimated by dividing
the detector into two independent sub-arrays ("odd pads" and "even pads")
and comparing the two reconstructed shower directions.
These two sub-arrays overlap spatially so that they are affected by the same
shower curvature.
Events with total number of hits $N_{hit}$ have been selected according to the
constraint $N_{odd}\simeq N_{even}\simeq N_{hit}/2$.
The distribution of the even-odd angle difference $\Delta \theta_{eo}$ is
shown in Fig. \ref{evenodd} for events in different multiplicity ranges 
and $\theta < 55^{\circ}$.
These distributions narrow, as expected, with increasing shower size.

\begin{figure}[htb]
\mbox{\epsfig{file=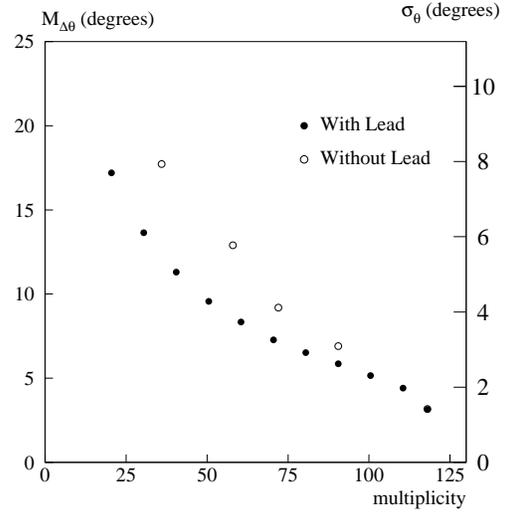,height=7.cm,width=7.cm}}
\caption{Median of $\Delta \theta_{eo}$ distribution as a function
of pad multiplicity.}
\label{median}
\end{figure}

The effect of the lead sheet on the angular resolution can be appreciated
in Fig. \ref{median}, where the median $M_{\Delta \theta_{eo}}$ of the 
distribution of
$\Delta \theta_{eo}$ as a function of pad multiplicity, for showers
reconstructed before and after the lead was added, is shown.
The improvement of the angular resolution is a factor $\sim$ 1.4 for 
$N_{hit}=50$ and decreases with increasing multiplicity.
Assuming that the angular resolution function for the entire array is Gaussian,
its standard deviation $\sigma$ is given by \cite{alexand92}
$\sigma_{\theta}={ {M_{\Delta \theta_{eo}}} \over {1.177 \times 2} }.$
This chessboard method yields a standard deviation $\sigma_{\theta}\sim
2.1^{\circ}$ for events with a pad multiplicity $\geq 100$.

\section{CONCLUSIONS}

A carpet of $\sim$ 50 $m^2$ of RPCs has been put in operation at the 
Yangbajing Laboratory in order to study the high altitude
performance of RPCs and the detector capability of imaging with high
granularity a small portion of the EAS disc, in view of an enlarged use
in Tibet (ARGO-YBJ experiment).

The results of this test confirm that RPCs can be operated efficiently 
($\geq 95\%$) to sample air showers at high altitude with excellent 
time resolution ($\sim 1$ $ns$). The analysis of data collected with a 
shower trigger suggest that the RPCs carpet capability of reconstructing 
the shower features is consistent with expectations. As a conclusion, the 
overall results of the test look well promising for future operation of the 
full detector.


\begin{thebibliography}{9}
\bibitem{abbr96} Abbrescia M. et al., {\it Astroparticle Physics with ARGO}, 
Proposal (1996). \\
Download this document at the URL: 
www1.na.infn.it/wsubnucl/cosm/argo/argo.html
\bibitem{bacci99} Bacci C. et al., submitted to Nucl. Instr. Meth. (1999).
\bibitem{bennet61} Bennet S. et al., J. Phys. Soc. of Japan, Vol. 17 Supp. 
A-III (1961) 196.
\bibitem{greisen56} Greisen K., Progr. Cosmic Ray Physics, Vol. III (1956).
\bibitem{alexand92} Alexandreas D.E. et al., Nucl. Instr. Meth. A311 (1992) 350.
\end{thebibliography}
\end{document}